\documentclass[fleqn,usenatbib,useAMS]{mnras}

\usepackage{graphicx}
\usepackage{subfigure}
\usepackage{enumerate}
\usepackage{amsmath}
\usepackage{appendix}

\usepackage{mathtools}

\usepackage{cleveref}

\usepackage{booktabs}
\usepackage{multirow}

\usepackage{xspace}

\newcommand{\dd}{\mathbf{d}}
\newcommand{\Epz}{$E_\text{p,z}$\xspace}
\newcommand{\Lpiso}{$L_\text{p,iso,52}$\xspace}
\newcommand{\Gf}{$\Gamma_\text{f}$\xspace}

\usepackage[T1]{fontenc}
\usepackage{ae,aecompl}

\usepackage{newtxtext,newtxmath}

\title[Empirical Correlation in ICMART Model]{Gamma-ray Burst Empirical Correlation between Peak Luminosity and Peak Energy in The ICMART Model}

\author[Shao \& Gao]{
Xueying Shao,$^{1, 2}$
He Gao,$^{1, 2}$\thanks{E-mail: gaohe@bnu.edu.cn}
\\
$^{1}$Institute for Frontier in Astronomy and Astrophysics, Beijing Normal University, Beijing 102206, China\\
$^{2}$Department of Astronomy, Beijing Normal University, Beijing 100875, China}

\date{}

\pubyear{2024}

\begin{document}
\label{firstpage}
\pagerange{\pageref{firstpage}--\pageref{lastpage}}
\maketitle

\begin{abstract}
    Internal-Collision-induced Magnetic Reconnection and Turbulence (ICMART) model is a widely accepted model for explaining how high-magnetization jets produce gamma-ray burst (GRB) prompt emissions.
    In previous works, we show that this model can produce:
    1) light curves with a superposition of fast and slow components;
    2) a Band-shaped spectrum whose parameters could follow the typical distribution of GRB observations;
    3) both ``hard to soft" and ``intensity tracking" patterns of spectral evolution.
    In this work, through simulations of a large sample with methods established in previous work, we show that the ICMART model can also explain the observed empirical relationships (here we focus on the Yonetoku and Liang relations), as long as the magnetic field strength in the magnetic reconnection radiation region is proportional to the mass of the bulk shell, and inversely proportional to the initial magnetization factor of the bulk shell.
    Our results suggest that during extreme relativistic magnetic reconnection events, an increase in magnetic field strength leads to more intense dissipation, ultimately resulting in a weaker residual magnetic field.
\end{abstract}


\section{Introduction}
\label{sec:introduction}
Gamma-ray bursts are the most luminous explosions in the universe.
Their bursty emission in the hard-X-ray/soft-$\gamma$-ray band is commonly referred as the ``prompt emission" \citep*{Zhang2018book}.
After decades of investigation the origin of the prompt emission still remain ambiguous and the controversy centers on the composition of GRB flow that power the prompt emission is matter-dominated or Poynting-flux-dominated.
The most representative model of the matter-dominated scenario is the internal shock model \citep{Rees1994}.
In this model, the erratic activity of the central engine and the angular spreading time of shells at different radii is believed to cause the temporal variability and the variable timescales of the light curve \citep{Rees1994,Kobayashi1997}.
A non-thermal spectrum is naturally generated as long as the collision radius is beyond the photosphere radius.
Although the internal shock model explains the origin of the prompt emission, it is difficult to account for the superposition of slow and fast components \citep{Gao2012} unless the central engine itself carries these two variability components in its time history of jet launching, which has no proper explanation in physics \citep{Hascoet2012}.
It also has difficulty on interpreting the hard-to-soft pattern between $E_\text{p}$ and flux because the peak energy and the peak luminosity are both determined by the crash intensity, which generate the tracking pattern naturally.
In addition, the internal shock model also faces other challenges \citep[][for a review]{Zhang2011},
including the efficiency problem \citep{Panaitescu1999,Kumar1999}, fast cooling problem \citep{Ghisellini2000,Kumar2008}, the electron number excess problem \citep{Daigne1998,Shen2009}, the missing bright photosphere problem \citep{Zhang2009,Daigne2002} and so on.

Poynting-flux-dominated models are proposed instead in order to solve these problems, which assume magnetic dissipation rather than kinetic dissipation happen to power the prompt emission through different mechanisms.
These kinds of models include MHD-condition-broken scenarios \citep{Usov1994,Zhang2002}, radiation-dragged dissipation model \citep{Meszaros1997}, slow dissipation model \citep{Thompson1994,Drenkhahn2002a,Giannios2008,Beniamini2017}, current-driven instabilities \citep{Lyutikov2003}, and forced magnetic reconnection \citep{Zhang2011,McKinney2012,Lazarian2019}.
In this work, we focus on the Internal-Collision-induced MAgnetic Reconnection and Turbulence (ICMART) model \citep{Zhang2011}, which supposes magnetic field distortion instead of shocks is yielded through mechanism collisions because the outflow contains a strong magnetic field with magnetization factor $\sigma_0 > 1$ that prevents the generation of shocks.
Significant energy has been proved to dissipate by a global simulation  when two highly magnetized blobs collide \citep{Deng2015}.
A Monte Carlo simulation which tracks the detailed evolution of the ICMART event and mimic an observed-like light curve shows that the fast component of the light curve results from the different luminosity, latitude and direction angle of each mini-jet contained in the ICMART event while the exponential growth of the mini-jets number and curvature effect constitute the slow component together \citep{Zhang2014}.
\citet{Shao2022} demonstrates that a Band-function-like spectrum with reasonable indexes can be produced by the synchrotron radiation happened in a decaying radiation magnetic field.
A ``hard to soft" pattern of $E_\text{p}$ is also naturally generated by the decaying background field due to the bulk expansion in this scenario while an ``intensity tracking" pattern is able to be obtained by superposition of a series of different ICMART events as long as they occur in specific orders \citep{Lu2012b,Shao2022}.

Besides the light curve and spectrum properties mentioned above, several empirical relations among observation parameters of the prompt emission are found in the last few decades.
For instance, \cite{Yonetoku2004} discovered a correlation between the GRB isotropic, bolometric peak luminosity ($L_{\rm p,iso}$) and the rest-frame peak energy ($E_{\rm p,z}$), which has a rough positive dependence $L_{\rm p,iso}\propto E_{\rm p,z}^{1.6}$ (henceforth Yonetoku relation).
Later, \cite{Liang} found a tighter relation between $L_{\rm p,iso}$ and $E_{\rm p,z}$ by involving the initial Lorentz factor ($\Gamma_0$) as the third parameter, $L_{\rm p,iso}\propto E_{\rm p,z}^{1.34}\Gamma_0^{1.32}$ (henceforth Liang relation).

This work will continue to use the Monte Carlo simulation method established in \cite{Shao2022} to test whether the ICMART model can successfully explain the Yonetoku and Liang relations through simulations of a large sample, and reveal the constraints these statistical relationships place on the relativistic magnetic dissipation process.

\section{Simulation}
\label{sec:ICMART model}

\subsection{Method}
\begin{figure}
    \centering
    \includegraphics[width=0.3\textwidth]{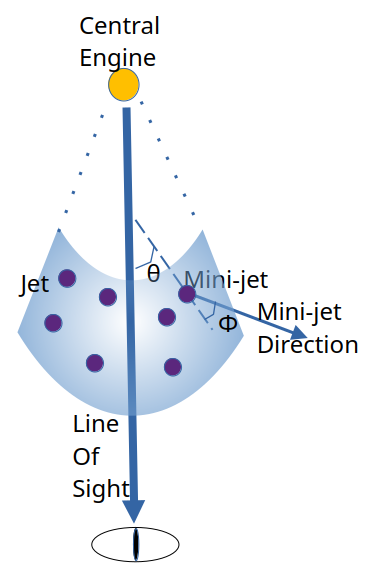}
    \caption{A schematic illustration about the spatial geometric relation between jet and mini-jets. The yellow circle is the central engine and the blue sector is the jet ejected from the engine. Each purple dot represent a mini-jet launched by the magnetic reconnection.
    }
    \label{fig:angle draft}
\end{figure}

Within the ICMART scenario, a succession of discrete and highly magnetized shells with variable mass, velocity and luminosity would be launched from the central engine.
Mechanical collisions between the late, fast shell and the early, slow shell would distort the magnetic field configurations and induce MHD turbulence.
Numerical simulations have demonstrated that the reconnection-driven magnetized turbulence has the ability to generate additional reconnections \citep{Takamoto2015,Takamoto2018,Kowal2017}, suggesting that the count of magnetic reconnections would increase exponentially.
For each reconnection event, the dissipated magnetic energy is approximately equal to $(\sigma_0 - 1) / \sigma_0$ times the magnetic energy within the reconnection region.
Here we assume that for each reconnection event, the dissipated magnetic energy is initially used to eject a bipolar minijet with Lorentz factor $\gamma \approx \sqrt{1+\sigma_0}$\citep{Zhang2014}.
We assume that about $f_\text{e}$ of the electrons within the reconnection region will be accelerated by Fermi acceleration and carry away about $10\%$ of the dissipated magnetic energy \citep{Drenkhahn2002a,Spitkovsky2008}, which will be transferred to photons later through synchrotron radiation to energize the prompt emission.
Other dissipated energy is used to increase the kinetic energy of the bulk shell through collisions between minijets with the matter surrounding the reconnection \citep{Drenkhahn2002a}.

Our simulation starts from a merged shell moving with mass $M_\text{bulk}$, Lorentz factor $\Gamma_0$, magnetized factor $\sigma_0 \equiv E_\text{m,0}/E_\text{k,0}$ ($E_\text{m,0}$ marks the initial magnetic energy and $E_\text{k,0}$ marks the initial kinetic energy) and radius $R_0$ when the reconnection cascade is triggered.
Three rest frames need to be invoked in the simulation: (1) the rest frame of the jet bulk, parameters in which are described with a single prime ($'$); (2) the rest frame of the mini-jet, parameters in which are described with double primes ($''$); (3) the rest frame of the observer (here we ignore the effect of cosmological expansion so the lab frame is stationary with respect to the observer frame), parameters in which are described with no prime.
Quantities in this three frames are connected with each other through two Doppler factors, i.e.,
\begin{eqnarray}
    \mathcal{D}_1 &= [\Gamma(1-\beta_\text{bulk}\cos{\theta})]^{-1},   \\
    \mathcal{D}_2 &= [\gamma(1-\beta_{jet}\cos{\phi})]^{-1},
\end{eqnarray}
where $\gamma$ is the relative Lorentz factor of mini-jet with respect to the jet bulk \citep{Zhang2014}, $\beta_{\rm bulk}$ and $\beta_{\rm jet}$ are the corresponding dimensionless velocities with respect to $\Gamma$ and $\gamma$. $\theta$ is the latitude of the mini-jet, i.e., the angle between the line of sight and the direction of the bulk at the location of the mini-jet and $\phi$ is the angle between the direction of the mini-jet and the direction of the bulk shell in the bulk comoving frame, a schematic illustration is presented in \cref{fig:angle draft}.

In the rest frame of the mini-jet, the accelerated electrons distribute in a power-law function read as
\begin{equation}
    Q(\gamma_e) = Q_0(\frac{\gamma_e}{\gamma_\text{e,m}})^{-p}, \gamma_\text{e,m}<\gamma_e<\gamma_\text{e,M},
    \label{eq:electron number distribution}
\end{equation}
where $Q_0$ is a normalization factor, $\gamma_\text{e,m}$ and $\gamma_\text{e,M} \equiv \sqrt{\frac{6\pi q_e}{\sigma_TB''_\text{e}}}$ are the minimum and maximum injected electron Lorentz factor, $B''_\text{e}$ is the magnetic field strength in the emission region in the jet frame.
Lacking full numerical simulations of magnetic turbulence and reconnection, here we introduce a free parameter $k$ to connect $B''_\text{e}$ with the bulk magnetic field strength as $B''_\text{e, 0} = \sqrt{k}B'/\gamma$. Moreover, we assume $B''_\text{e}$ decays with the radius of the bulk as
\begin{equation}
    B''_\text{e}(R) = B''_\text{e,0}(\frac{R}{R_0})^{-b},
\end{equation}
where b would be much larger than 1, due to the rapid consumption of magnetic energy in the reconnection process. The bulk magnetic field strength decreases as the jet expanding, which reads
\begin{equation}
    B'(R) = B'_0(\frac{R}{R_0})^{-1},
    \label{eq:decaying magnetic strength}
\end{equation}
where $B'_0 = \sqrt{8\pi\sigma_0\Gamma_0Mc^2 / V'_0}$ is the initial magnetic field strength and $V'_0$ is the initial volume of the bulk \citep{Shao2022}.

For the ith magnetic reconnection $Q_0$ and $\gamma_\text{e, m}$ can be achieved by solving the electron number and energy conservation equation, which are written as \citep{Shao2022}
\begin{equation}
    Q_0 (t''_\text{e}-t''_\text{i})\int_{\gamma_\text{e,m}}^{\gamma_\text{e,M}}(\frac{\gamma_e}{\gamma_\text{e,m}})^{-p}\dd \gamma_e = f_\text{e}\int_{t'_\text{i}}^{t'_\text{e}}{S'}^2 v'_\text{in} n'_\text{e}\dd t',\\
    \label{eq:particle number conservation}
\end{equation}
and
\begin{equation}
    \frac12 \delta E'_\text{m, i} = \int_{t''_\text{i}}^{t''_\text{e}}\int_{\gamma_\text{e,m}}^{\gamma_\text{e,M}}Q_0(\frac{\gamma_e}{\gamma_\text{e,m}})^{-p}\dd \gamma_e \dd t''(\gamma_e-1)m_\text{e}c^2,
    \label{eq:particle energy conservation}
\end{equation}
where $t''_\text{i}$ and $t''_\text{e}=t''_\text{i}+T''_0(t''_i)$ are the starting and ending time of the magnetic reconnection,  $T_0''(t''_i)\equiv S''(t''_\text{i})/v''_\text{in}$ is the duration time of the reconnection, $n'_\text{e} \propto R^{-2}$ is the number density of electrons in the bulk frame, $S' = S'_0(\frac{R}{R_0})^s$ is the size of the magnetic reconnection and $v'_\text{in}$ is the inflow velocity of the magnetic field line.
$\delta E'_\text{m, i}$ is the dissipated energy for the magnetic reconnection, which could be estimated as
\begin{equation}
    \delta E'_\text{m, i} = \frac{(\sigma_0-1){B'}^2(t_i){S'}^3(t_i)}{8\pi\sigma_0}. \label{eq:dissipated energy of single reconnection}
\end{equation}

Given the electron distribution, the synchrotron radiation power in the rest frame of the mini-jet could be given by \citep{Rybicki1985}
\begin{equation}
\label{eq:pnup}
P''_{\nu''} = \frac{\sqrt{3} q_e^3 B''_\text{e}}{m_{\rm e} c^2}
	    \int_{\gamma_{\rm e,m}}^{\gamma_{\rm e,M}}
	    \left( \frac{dN_{\rm e}''}{d\gamma_{\rm e}} \right)
	    F\left(\frac{\nu ''}{\nu_{\rm cr}''} \right) d\gamma_{\rm e},
\end{equation}
where $q_e$ is electron charge and $\nu_{\rm cr}'' = 3 \gamma_{\rm e}^2 q_e B''_\text{e} / (4 \pi m_{\rm e} c)$ is the
characteristic frequency of an electron with Lorentz factor $\gamma_e$. In the observer frame, each mini-jet radiation corresponds to a single pulse starting from $t'_\text{i}/\mathcal{D}_1$ to $t'_\text{e}/\mathcal{D}_1$, with radiation power
\begin{equation}
    P_{\nu}= \mathcal{D}_1^3 \mathcal{D}_2^3 P''_{\nu''}.
\end{equation}
Here we use Gaussian shape to simulate each single pulse in the light curve. The total observed radiation power is the superposition of the emission from all these mini-emitters. For a given time interval in the observer frame ($t_1$, $t_2$), the time resolved spectrum is given by
\begin{equation}
    P_{\nu}(\nu) = \sum_{N_\text{tot}}\sum_{t=t_1}^{t=t_2} P_\nu(\nu, t),
\end{equation}
where $N_\text{tot}$ is the total count of magnetic reconnections. The ICMART process can efficiently dissipate a significant proportion of magnetic energy. Here, we use $f_\text{p}$ to denote the magnetic energy dissipation ratio of the entire reconnection cascade process. Therefore, the total number of magnetic reconnection events can be estimated through the following formula,
\begin{equation}
    f_\text{p}\sigma_0\Gamma_0Mc^2 = \sum_{i=1}^{N_\text{tot}}\Gamma_i\delta E'_\text{m, i}. \label{eq:dissipated energy conservation}
\end{equation}
For one ICMART event, we assume an abrupt cessation of the cascade process, so that magnetic reconnection happen from $t_\text{p}-T_0$ to $t_\text{p}$ would affect the observation at peak time ($t_\text{p}$). The light curve after $t_\text{p}$ is therefore contributed by the high-latitude emission from other mini-jets not along the line of sight due to the “curvature effect” delay.

\subsection{Initial Conditions}

For each simulation, the initial conditions include five independent parameters: the mass of the bulk shell $M$, the magnetized factor $\sigma_0$, the fraction of the dissipated magnetic energy $f_\text{p}$, the fraction of electrons accelerated by Fermi acceleration of each magnetic reconnection $f_\text{e}$ and the ratio between the radiated magnetic field strength and the bulk magnetic field strength $k$.
Previous studies have shown that to generate a spectrum consistent with GRB observational data, there are certain requirements for these physical parameters of ICMART events\citep{Uhm2015,LucasUhm2018,Shao2022}. For instance, the magnetic field in the radiation region cannot be too strong or too weak, as this would cause the peak of the radiation spectrum generated by ICMART events to deviate from gamma ray band or the shape of the radiation spectrum to deviate from the Band spectrum type. From a natural perspective, ICMART events with different parameter combinations should all have the potential to occur. Those events that fail to produce GRBs may correspond to other astronomical phenomena, such as soft X-ray flashes or Fast X-ray transients.

We construct one simulation by selecting random values from the parameter value space of these five parameters. Here we take a uniform distribution in logarithmic space for $M$ and $k$ in the ranges of $(10^{23}, 10^{27})\text{g}$ and $(10^{-12}, 10^{-2})$, respectively. We take a uniform distribution in linear space for $\sigma_0$ and $f_\text{e}$ in the ranges of $(12.5, 50)$ and $(0.3, 0.5)$, respectively. These parameters are set as a continuation of \cite{Shao2022}, in which we found their combinations are easier to generate a validate GRB event. For the setting of $f_\text{p}$, we referred to the observed distribution of radiation efficiencies of GRBs in the gamma-ray band \citep{Zhang2007}, adopting a uniform distribution within the range of $(0.01, 0.5)$.

In an ICMART event, each magnetic reconnection has a distinct position and orientation. We consider the latitude of the mini-jet with respect to the line of sight $\theta$ within $(0, 2/\Gamma_0)$ and the angle between the mini-jet direction and radial direction of the bulk within the bulk comoving frame $\phi$ within $(0, \pi/2)$ for each reconnection. Additionally, we assume that the initial size $S'_0$ and the injecting velocity of the magnetic field line $v'_\text{in}$ of the magnetic reconnection are $6\times10^{10} \text{cm}$ and $1.5\times10^{9} \text{cm/s}$, respectively; the decaying index of the radiation magnetic field $b$ is set to $30$ and the growth index the size of the magnetic reconnection $s$ is set to $0.2$. These parameters are all set as a continuation of \cite{Shao2022}. As for the bulk shell, the initial radius $R_0$ is set to $10^{14} \text{cm}$.

\subsection{Results}

\begin{figure}
    \centering
    \subfigure[]{
    \includegraphics[width=0.3\textwidth]{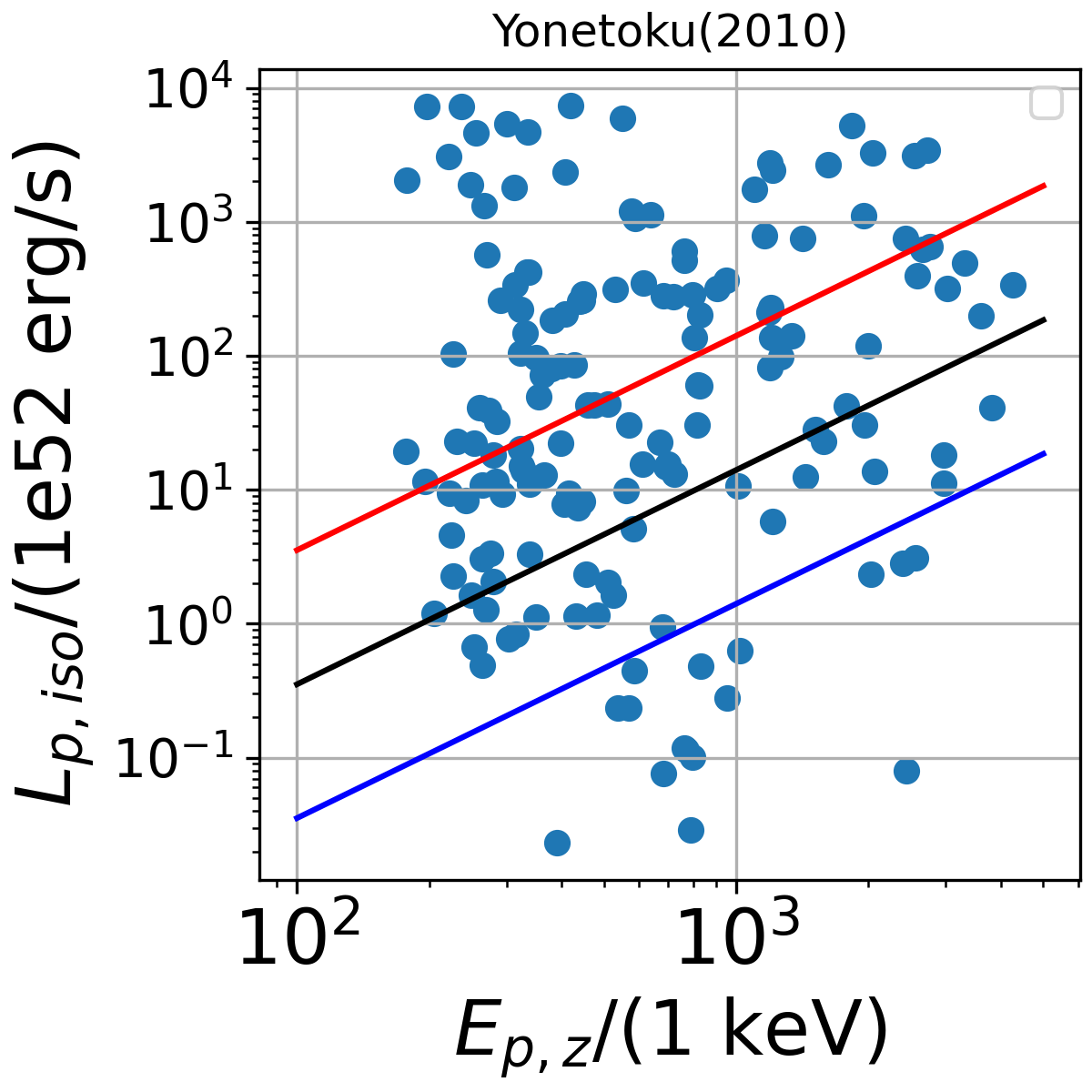}
    }
    \subfigure[]{
    \includegraphics[width=0.3\textwidth]{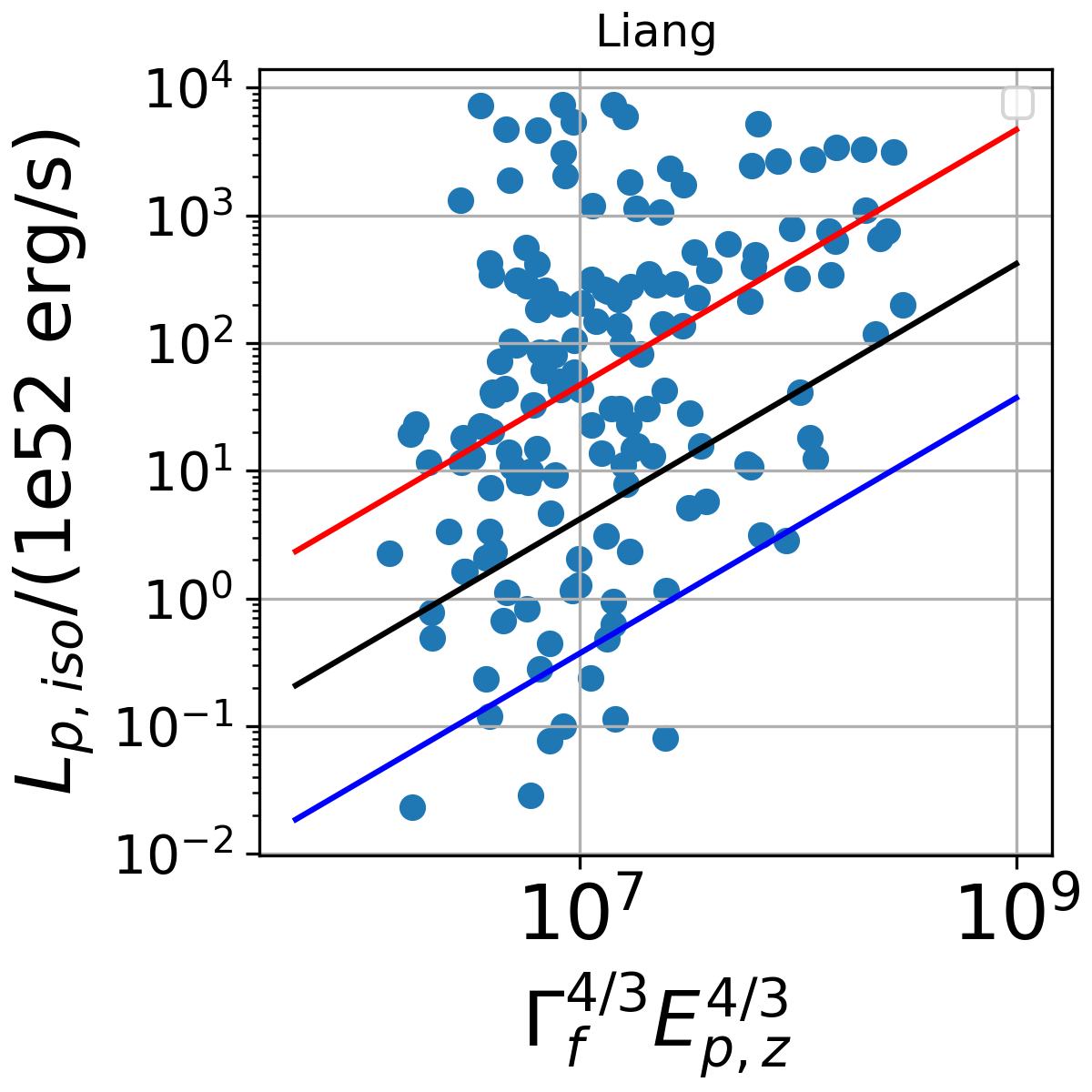}
    }
    \caption{Each blue dot represents a simulated GRB.
    The black lines represent: (a) the Yonetoku relation; (b) the Liang relation, with the red and blue lines delineating their $\pm 3\sigma$ error.
    }
    \label{fig:empirical relations}
\end{figure}

Provided above initial conditions, we simulate 2000 ICMART events.
According to the observation features of GRBs, we selected the simulated events that meet the following three conditions: 1) $100 < \Gamma_\text{f}< 1000$ \citep{Racusin2011} , $\Gamma_\text{f}$ is the Lorentz factor of the bulk at the end of the reconnections cascade; 2) the simulated spectrum is Band shape, the criterion is whether the R-square value between the simulated spectrum and the fitted Band function is bigger than $0.96$ \footnote{In this work, we mainly compare the simulation results with empirically derived relationships obtained through observations. The $E_p$ data used in deriving empirical relationships such as the Yonetoku/Liang relation are typically obtained by fitting with the Band function. Therefore, we also choose to fit the simulation results using the Band function here. We have also tested the use of the cutoff power law model and found that it does not affect our conclusions.}; 3) the peak energy of the spectrum is within $5 {\rm keV} < E_\text{p} < 5000 {\rm keV}$ \citep{Yonetoku2010}, $E_\text{p}$ corresponds to the integral energy of the number of photons arriving within $\sim 1$ second in the lab frame and is measured as the peak energy of the fitting Band function.
Eventually, we obtained 157 ICMART events that can be considered as successfully producing GRBs.
For each of the selected events, we recorded their isotropic peak luminosity \Lpiso, the rest frame peak energy \Epz, and the Lorentz factor of the bulk shell at the end of the reconnection cascade process \Gf.

Two empirical relations considered in this paper are the Yonetoku relation reads as \citep{Yonetoku2010}
\begin{equation}
\frac{L_\text{p,iso}}{10^{52}\text{erg}\cdot\text{s}^{-1}} = 10^{0.43\pm0.037}\times[\frac{E_\text{p,z}}{355\text{keV}}]^{1.60\pm0.082},
\label{eq:Yonetoku}
\end{equation}
and the Liang relation reads as \citep{Liang}
\begin{equation}
\frac{L_\text{p,iso}}{10^{52}\text{erg}\cdot\text{s}^{-1}} = 10^{-6.38\pm0.35}(\frac{E_\text{p,z}}{1\text{keV}})^{1.34\pm0.14}\Gamma_\text{f}^{1.32\pm0.19}.
\label{eq:Liang}
\end{equation}

We plotted the selected simulated samples and the two empirical relations in \cref{fig:empirical relations}, and found that only about half of the samples fall within the $3\sigma$ range of both relations.
This indicates that even if ICMART events can produce individual GRB samples that meet the observational requirements independently, it is not possible to statistically satisfy the observed statistical correlation if the initial conditions are randomly selected without any constraints between parameters.

\section{Parameter relationship}
\begin{figure}
    \centering
    \includegraphics[width=0.45\textwidth]{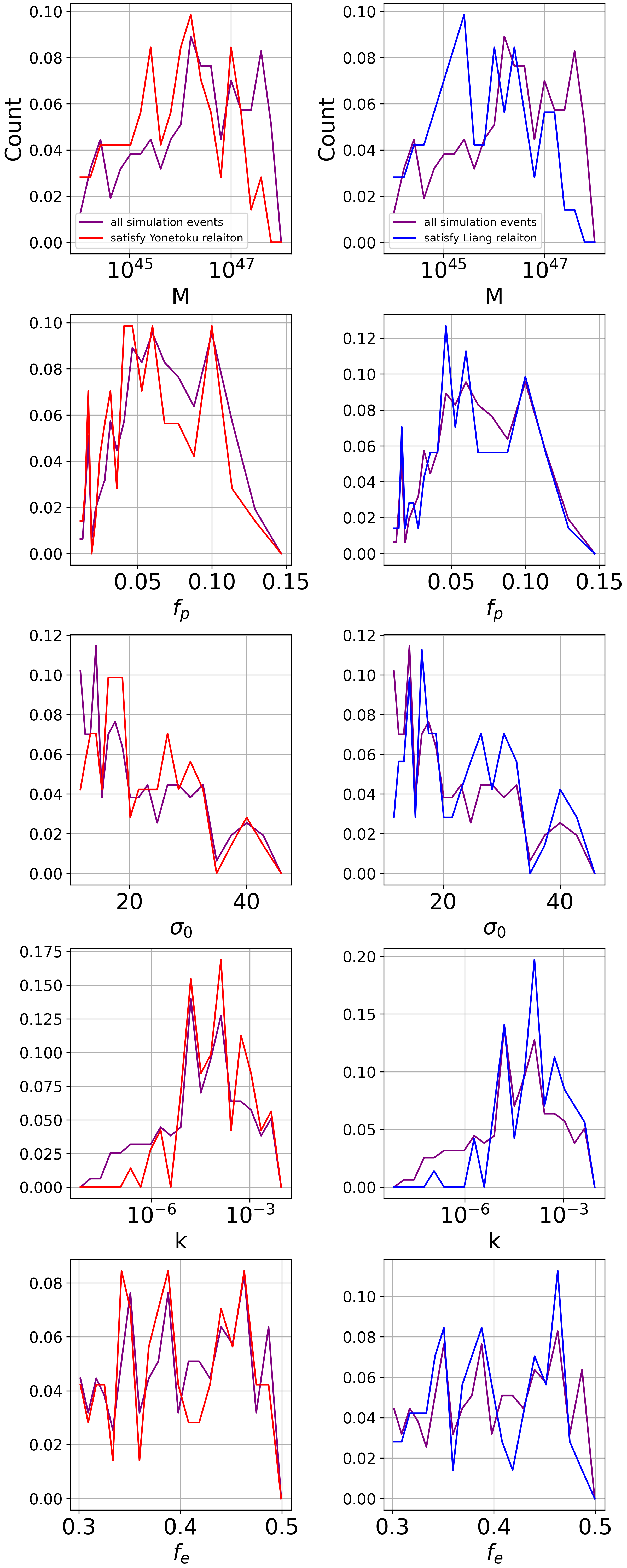}
    \caption{Distributions of initial setup parameters.
    The purple lines are for the total 157 events.
    The red and blue lines are for samples that satisfy the Yonetoku relation and the Liang relation separately.
    }
    \label{fig:variables distribution}
\end{figure}
\begin{figure}
    \centering
    \includegraphics[width=0.45\textwidth]{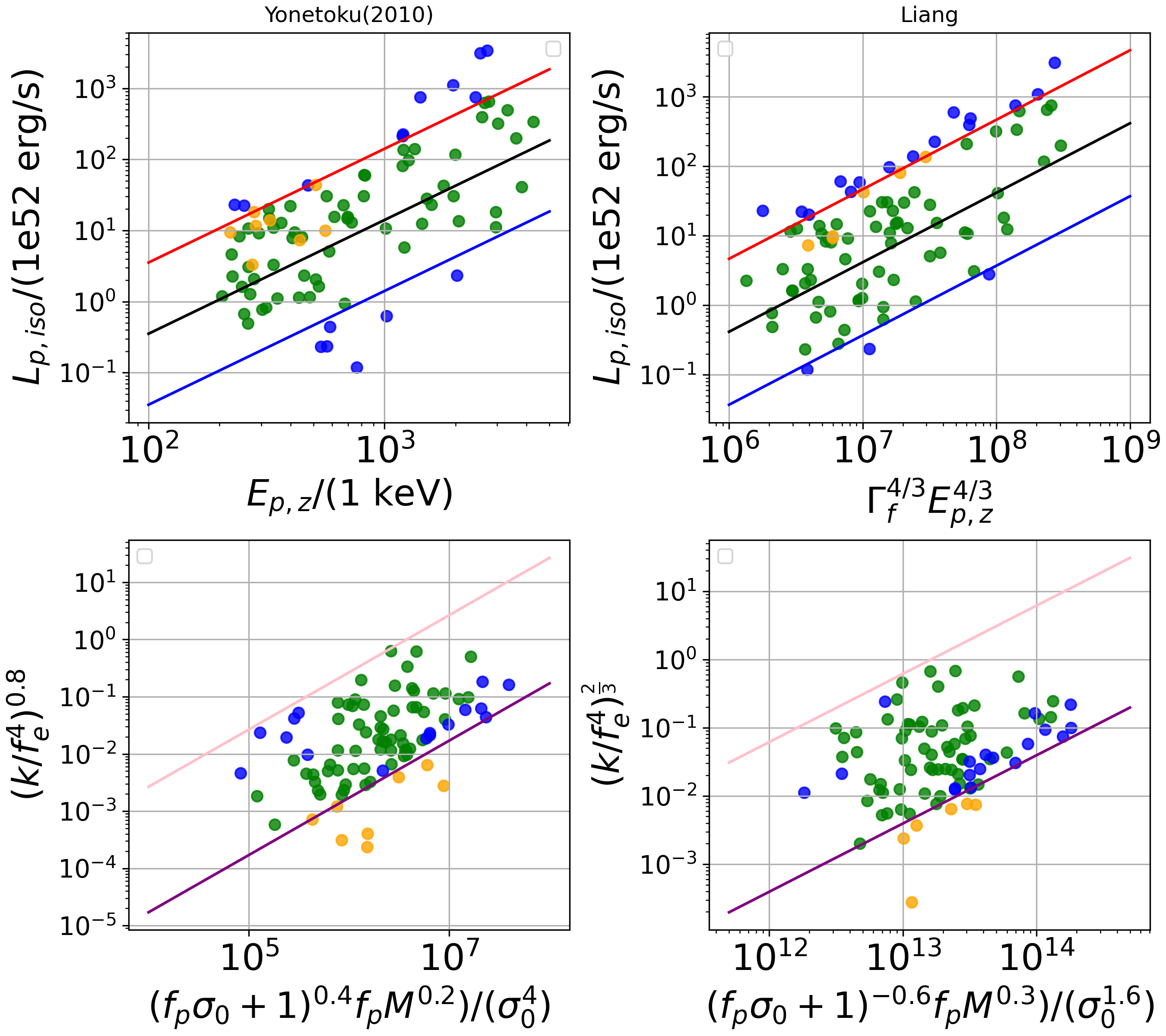}
    \caption{This figure shows the correspondence between the correlation relationship among the ICMART model parameters given by \cref{eq:k-fe relation} and the observed empirical relationship.
    Dots with different colors represent different kinds of events: (1) green: satisfy both the empirical relations and \cref{eq:k-fe relation}; (2) blue: do not satisfy the empirical relations but satisfy \cref{eq:k-fe relation}; (3) orange: satisfy the empirical relations but do not satisfy \cref{eq:k-fe relation}.
    In the top panel, the black lines represent the two empirical relations and the red and blue lines mark their $\pm3\sigma$ error range.
    In the bottom panel, the pink and the purple lines delineate the upper and lower limits of \cref{eq:k-fe relation}.
    }
    \label{fig:k-fe relation}
\end{figure}
\begin{figure}
    \centering
    \includegraphics[width=0.45\textwidth]{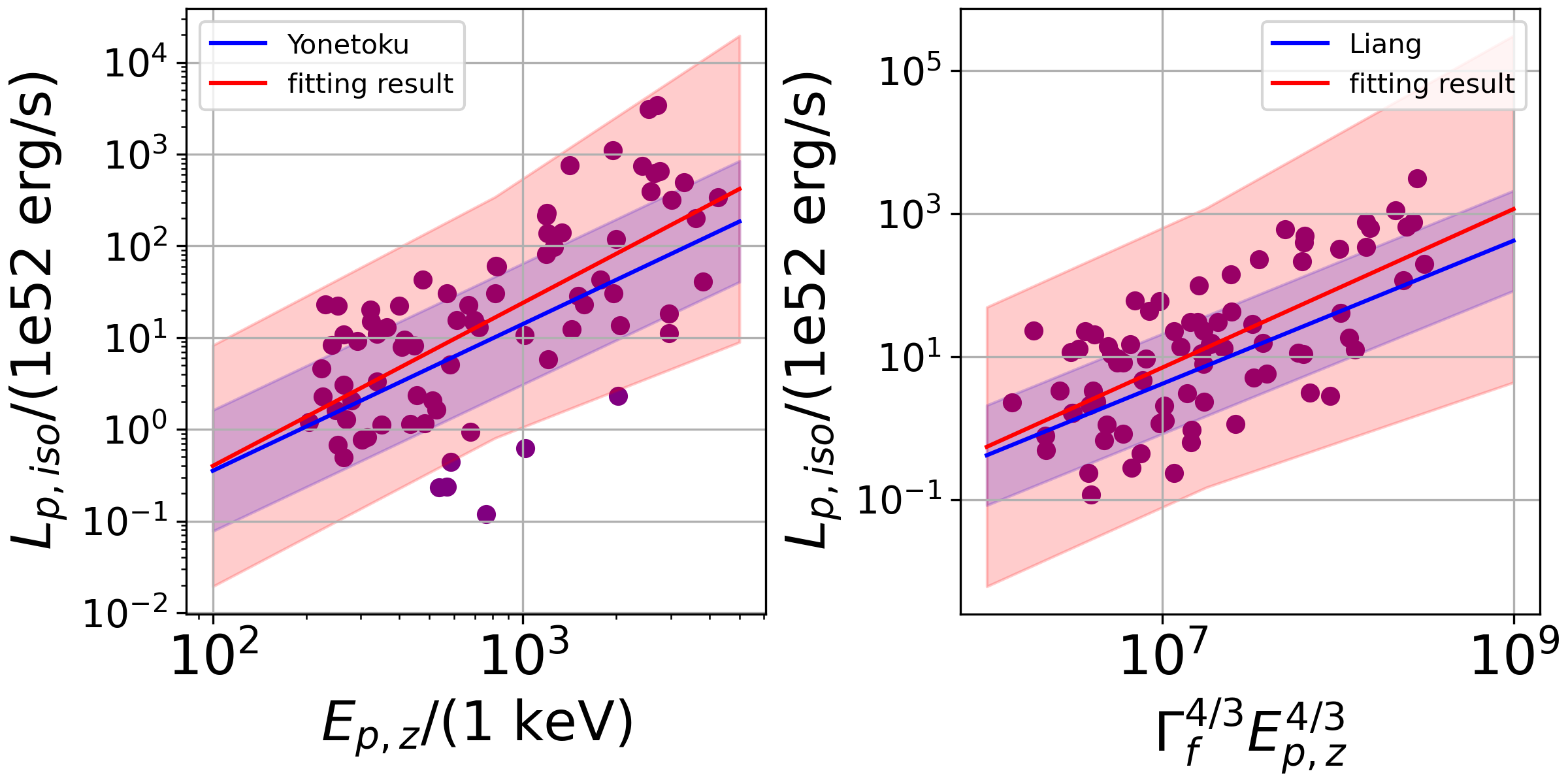}
    \caption{Each purple dot in this figure represents an ICMART event whose parameters satisfy \cref{eq:k-fe relation}.
    The red lines are the best fit results for these simulated events, with the shades delineating their $\pm2\sigma$ region.
    The blue lines show the the statistical relationship obtained from the observational data, with the shades delineating their $\pm2\sigma$ region.
    }
    \label{fig:fit_deduce_data}
\end{figure}

In \cref{fig:variables distribution}, we plot the distributions of initial setup parameters for both the total 157 events and the samples that satisfy the two empirical relations.
We find that the distributions of the former sample do not show any special features compared to the latter.
This result indicates that the emergence of empirical relations should not be attributed to the special distribution of certain initial parameters but rather likely due to the correlation among parameters.

In the following, based on the ICMART model theory, we first use a semi-analytical approach to derive the possible correlation among parameters to make GRBs satisfy the Yonetoku relation and the Liang relation.
Then, we test the derived results with our simulated samples.

The Yonetoku relation and the Liang relation can be generalized as
\begin{equation}
    L_\text{p,iso} \propto E_\text{p,z}^{2\alpha}\Gamma_\text{f}^{2\beta},
    \label{eq:empirical relation}
\end{equation}
with $\alpha = 0.8$ and $\beta = 0$ for the Yonetoku relation and $\alpha \approx \beta \approx \frac23$ for the Liang relation.
In our simulation, the recorded luminosity is beaming corrected, so that the isotropic peak luminosity could be estimated as
\begin{equation}
L_\text{p, iso} = \frac{4\pi L_\text{p}}{\Delta \Omega},
\end{equation}
where $L_\text{p}$ is the peak luminosity of the simulated light curve, and $\Delta\Omega$ represents the solid angle of the jet \citep{Zhang2018book}.
In the simulation, we assumed that the magnetic reconnections satisfy an exponential growth and the number doubles after each generation.
The cascade stops after consuming a certain proportion of magnetic energy.
Therefore, the peak luminosity in the simulation essentially corresponds to the radiation contribution from the last generation of magnetic reconnection, namely
\begin{equation}
    L_\text{p} = \frac{0.1\Delta E_\text{m, f}}{T_\text{0,f}},\label{eq:Lp}
\end{equation}
where $T_\text{0,f}\sim 0.1s$ and $\Delta E_\text{m, f}$ are the timescale and dissipated magnetic energy of magnetic reconnections of the last generation.
Here $10\%$ of the dissipated magnetic energy is assumed to be transferred into radiation.
Henceforth we use subscript ($\text{f}$) to denote the physical quantities corresponding to the last generation of magnetic reconnection.
$\Delta E_\text{m, f}$ can be estimated as
\begin{equation}
    \Delta E_\text{m, f} = \Gamma_\text{f}N_\text{f}\delta E'_\text{m, f}, \label{eq:dissipated magnetic energy}
\end{equation}
where $\delta E'_\text{m, f}$ is the dissipated magnetic energy of a single reconnection and $N_\text{f}$ is the number of the reconnections.
$N_\text{f}$ can be calculated as
\begin{equation}
    N_\text{f} = \frac12N_\text{tot} = \frac12\int_0^{t_\text{p}}e^{\kappa t} \dd t, \label{eq:Ntot}
\end{equation}
where $t_\text{p}$ is the end time of reconnection cascade and $\kappa\equiv\ln2/T_\text{0}$ is the growth index of the reconnection number.
Substituting \cref{eq:dissipated energy of single reconnection} and \cref{eq:Ntot} into \cref{eq:dissipated energy conservation} one can obtain
\begin{equation}
    \int_0^{t_p}e^{\kappa t}(\frac{R(t)}{R_0})^{-2+3s}\dd t \approx 2N_\text{f}\overline{(\frac{R(t)}{R_0})^{-2+3s}} = \frac{8\pi f_\text{p}\sigma_0 M c^2}{{B'}_0^2{S'}_0^3},
\end{equation}
where $\sigma_0+1$ is approximated to $\sigma_0$.
So $N_\text{f}$ can be approximated as
\begin{equation}
    N_\text{f} \approx \frac{200\pi f_\text{p}R_0^3}{{S'}_0^3\Gamma_0^3}, \label{eq:Nf}
\end{equation}
where $\overline{(R(t)/R_0)^{-2+3s}}\sim0.01$ for our selected parameters.

The kinetic energy conservation of the bulk shell reads as
\begin{equation}
    \Gamma_\text{f}M c^2 - \Gamma_0Mc^2 = 0.9f_\text{p}\sigma_0\Gamma_0Mc^2, \label{eq:kinetic change}
\end{equation}
where $90\%$ of the dissipated magnetic energy is assumed to be transferred to the bulk movement.
So that
\begin{equation}
    \Gamma_\text{f} = \Gamma_0(1 + 0.9f_\text{p}\sigma_0). \label{eq:Gf}
\end{equation}
Substituting \cref{eq:dissipated energy of single reconnection}, \cref{eq:Nf}, \cref{eq:Gf} and \cref{eq:dissipated magnetic energy} into \cref{eq:Lp} one can obtain
\begin{equation}
    L_\text{p} = \frac{5(\sigma_0-1)(1+0.9f_\text{p}\sigma_0){B'}_0^2f_\text{p}R_0^3}{4T_\text{0,f}\sigma_0\Gamma_0^2}(\frac{R_\text{f}}{R_0})^{-2+3l}
\end{equation}

Corresponding to the peak luminosity, the peak energy of the spectrum should be primarily determined by the energy spectrum peak of the mini-jets pointing in the direction of the line of sight.
In the simulation, the spectrum of a single magnetic reconnection is taken as the spectrum of synchrotron radiation in a decaying magnetic field \citep{Uhm2013,Shao2022}, so the peak energy can be estimated as
\begin{equation}
    E_\text{p,z} = \frac{3q\Gamma_\text{f}\gamma}{4\pi m_\text{e} c}{B''}_\text{e} {\gamma''}^2_\text{e, m}.
\end{equation}

Based on the above results of formula derivation, we found that under the framework of ICMART model, the initial parameters of the model need to satisfy the following relationship in order to make the empirical relationship described in \cref{eq:Yonetoku,eq:Liang} hold:
\begin{equation}
    (\frac{k}{f_\text{e}^4})^\alpha = F_K\frac{(f_\text{p}\sigma_0+2)^{3-2\alpha-2\beta}M^{1-\alpha}}{\sigma_0^{4-2\beta}}, \label{eq:k-fe relation}
\end{equation}
where $F_K$ is a factor that is given as
\begin{equation}
    F_K = \frac{200\pi c^{2-2\alpha} C_P v'_\text{in}R_\text{f}^{-2+2l+2\alpha}}{2^{2-2\alpha-2\beta}C_R C_E^{2\alpha}C_I^{2\alpha}S'_0R_0^{-2+2l-\alpha}},\label{eq:FK}
\end{equation}
where $C_P = \frac{1}{\pi10^{52}}$, $C_E = \frac{3hq}{4\pi m_\text{e} c}$ and $C_I = \frac{1837(-p+2)}{-p+1}$.
$C_R$ is the coefficient of the empirical relation and taken as $2.23 \times 10^{-4\pm1}$ \citep{Yonetoku2010} for the Yonetoku relation and $10^{-6.38\pm1.05}$ \citep{Liang} for the Liang relation.

In the following, we will test the correspondence between the correlation relationship among the ICMART model parameters given by \cref{eq:k-fe relation} and the observed empirical relationship, by using our simulated samples.
The results are shown in \cref{fig:k-fe relation}.
The left panels in \cref{fig:k-fe relation} are results for the Yonetoku relation.
Firstly, we found that out of the $157$ samples we simulated, $45\%$ fell within the $3\sigma$ range of the Yonetoku relation.
Among these $71$ samples, $89\%$ followed the parameter relationship given by \cref{eq:k-fe relation}.
The upper and lower boundaries of the parameter relationship correspond to $F_K$ values obtained by traversing all combinations of initial parameters for the $157$ samples.
A few points that did not fit the parameter relationship were not significant outliers.
Conversely, we found that $50\%$ of the $157$ samples followed the parameter relationship given by \cref{eq:k-fe relation}.
Among these $79$ samples, $80\%$ fell within the $3\sigma$ range of the Yonetoku relation, while $16$ points that did not fall within the $3\sigma$ range were still close to the boundary.
The right panels in \cref{fig:k-fe relation} are results for the Liang relation.
Similarly, we found that out of the $157$ samples we simulated, $45\%$ fell within the 3 sigma range of the Liang relation.
Among these $70$ samples, $91\%$ followed the parameter relationship given by \cref{eq:k-fe relation}.
A few points that did not fit the parameter relationship were very close to the lower boundary of relationship.
Conversely, we found that $52\%$ of the $157$ samples followed the parameter relationship given by \cref{eq:k-fe relation}.
Among these $82$ samples, $78\%$ fell within the $3\sigma$ range of the Liang relation, while $18$ points that did not fall within the $3\sigma$ range were also very close to the boundary of the $3\sigma$ region.

Considering that we made some simplifying assumptions and ignored some randomness in the magnetic reconnection process during the derivation of \cref{eq:k-fe relation}, it is tolerable that a small number of samples do not simultaneously satisfy the parameter relationship and the observed empirical relationship.
On the other hand, we cannot rule out the possibility that the statistical relationships obtained from the existing observational data may have certain systematic errors due to insufficient sample size.
For example, \cite{Zitouni2016} took a sample of bright GRBs observed by Swift/BAT and found the sample to be in good agreement to the Yonetoku relation. However, \cite{Xu2023} pointed out that some relatively fainter GRBs may significantly deviate from the Yonetoku relation due to potential contamination from off-axis samples.

In \cref{fig:fit_deduce_data}, we fitted all the points in the simulated sample that satisfy the parameter relationship, and found that the fitting results are very similar to the statistical relationship obtained from the real data, with only a slightly larger error range.
Overall, our results are sufficient to show that the ICMART model can explain the observed empirical relationships given a certain special relationship between the model parameters.

In order to better understand the physical meaning of the parameter relationship, we can replace $k$ with the magnetic field strength in the emission region in the lab frame $B_\text{e}\propto \frac{\sqrt{k\sigma_0M\Gamma_0^3}}{\Gamma_\text{f}}$ and then have \footnote{Here, we consider $f_\text{e}$ as an independent parameter that is not related to other parameters. Moreover, since $(f_\text{p}\sigma_0+2)^{3-4\alpha-2\beta}$ changes within a small range of values, we also ignore its dependence on $\sigma_0$.}
\begin{equation}
    {B}_\text{e}^{2\alpha}\propto\frac{M}{\sigma_0^{4-2\beta}}.\label{eq:BM relation}
\end{equation}
For Yonetoku and Liang relations, we have ${B}_\text{e}^{1.6}\propto{M}/{\sigma_0^4}$ and ${B}_\text{e}^\frac43\propto{M}/{\sigma_0^\frac83}$, respectively.
The physical meaning of \cref{eq:BM relation} is relatively clear: in order to make GRBs generated by ICMART events satisfy the statistical Yonetoku and Liang relations, the magnetic field strength in the magnetic reconnection radiation region is required to be proportional to the mass of the bulk shell, and inversely proportional to the initial magnetization factor of the bulk shell.

\section{Discussion}
\label{sec:discussion}

Previous works have verified that the ICMART model can well explain the light curve features (such as the combination of fast and slow components), spectral features (such as the Band spectrum with a relatively high low-energy spectral index), and spectral evolution features (both ``hard to soft" pattern and ``intensity tracking" pattern) of gamma-ray bursts through simulating individual GRBs.

In this work, through simulations of a large sample with method established in \cite{Shao2022}, we show that the ICMART model can also explain the observed empirical relationships (here we focus on the Yonetoku and Liang relations), as long as the magnetic field strength in the magnetic reconnection radiation region is proportional to the mass of the bulk shell, and inversely proportional to the initial magnetization factor of the bulk shell, i.e., ${B}_\text{e}^{2\alpha}\propto\frac{M}{\sigma_0^{4-2\beta}}$.
Our results suggest that during extreme relativistic magnetic
reconnection events, local magnetic dissipation occurs more violently in the presence of a stronger magnetic field, resulting in a weaker residual magnetic field.
This can be seen as an observational insights from GRB data for the study of relativistic magnetic reconnection physics, which is yet to be tested through dedicated numerical simulations in the future.

This work did not involve empirical relationships such as the Amati relation \citep{Amati2006} that contain total radiation energy, mainly because total radiation energy is related to the total radiation timescale, which is determined by the nature of the central engine activity and is not directly related to jet composition and energy dissipation process.

\section*{Data Availability}

The data underlying this article will be shared on reasonable request to the corresponding author.

\section*{Acknowledgements}
    We thank the helpful discussion with Bing Zhang.
    This work is supported by the National Natural Science Foundation of China (Projects 12373040,12021003), the National SKA Program of China (2022SKA0130100) and the Fundamental Research Funds for the Central Universities.
\clearpage

\bibliographystyle{mnras}
\bibliography{paper.bib}

\begin{thebibliography}{}
\makeatletter
\relax
\def\mn@urlcharsother{\let\do\@makeother \do\$\do\&\do\#\do\^\do\_\do\%\do\~}
\def\mn@doi{\begingroup\mn@urlcharsother \@ifnextchar [ {\mn@doi@}
  {\mn@doi@[]}}
\def\mn@doi@[#1]#2{\def\@tempa{#1}\ifx\@tempa\@empty \href
  {http://dx.doi.org/#2} {doi:#2}\else \href {http://dx.doi.org/#2} {#1}\fi
  \endgroup}
\def\mn@eprint#1#2{\mn@eprint@#1:#2::\@nil}
\def\mn@eprint@arXiv#1{\href {http://arxiv.org/abs/#1} {{\tt arXiv:#1}}}
\def\mn@eprint@dblp#1{\href {http://dblp.uni-trier.de/rec/bibtex/#1.xml}
  {dblp:#1}}
\def\mn@eprint@#1:#2:#3:#4\@nil{\def\@tempa {#1}\def\@tempb {#2}\def\@tempc
  {#3}\ifx \@tempc \@empty \let \@tempc \@tempb \let \@tempb \@tempa \fi \ifx
  \@tempb \@empty \def\@tempb {arXiv}\fi \@ifundefined
  {mn@eprint@\@tempb}{\@tempb:\@tempc}{\expandafter \expandafter \csname
  mn@eprint@\@tempb\endcsname \expandafter{\@tempc}}}

\bibitem[\protect\citeauthoryear{Amati}{Amati}{2006}]{Amati2006}
Amati L.,  2006, \mn@doi [Monthly Notices of the Royal Astronomical Society]
  {10.1111/j.1365-2966.2006.10840.x}, 372

\bibitem[\protect\citeauthoryear{Beniamini \& Giannios}{Beniamini \&
  Giannios}{2017}]{Beniamini2017}
Beniamini P.,  Giannios D.,  2017, \mn@doi [{\textbackslash}mnras]
  {10.1093/mnras/stx717}, 468, 3202

\bibitem[\protect\citeauthoryear{Daigne \& Mochkovitch}{Daigne \&
  Mochkovitch}{1998}]{Daigne1998}
Daigne F.,  Mochkovitch R.,  1998, \mn@doi [Monthly Notices of the Royal
  Astronomical Society] {10.1046/j.1365-8711.1998.01305.x}, 296

\bibitem[\protect\citeauthoryear{Daigne \& Mochkovitch}{Daigne \&
  Mochkovitch}{2002}]{Daigne2002}
Daigne F.,  Mochkovitch R.,  2002, \mn@doi [Monthly Notices of the Royal
  Astronomical Society] {10.1046/j.1365-8711.2002.05875.x}, 336

\bibitem[\protect\citeauthoryear{Deng, Li, Zhang  \& Li}{Deng
  et~al.}{2015}]{Deng2015}
Deng W.,  Li H.,  Zhang B.,   Li S.,  2015, \mn@doi [Astrophysical Journal]
  {10.1088/0004-637X/805/2/163}, 805

\bibitem[\protect\citeauthoryear{Drenkhahn \& Spruit}{Drenkhahn \&
  Spruit}{2002}]{Drenkhahn2002a}
Drenkhahn G.,  Spruit H.~C.,  2002, \mn@doi [Astronomy and Astrophysics]
  {10.1051/0004-6361:20020839}, 391, 1141

\bibitem[\protect\citeauthoryear{Gao, Zhang  \& Zhang}{Gao
  et~al.}{2012}]{Gao2012}
Gao H.,  Zhang B.-B.,   Zhang B.,  2012, \mn@doi [The Astrophysical Journal]
  {10.1088/0004-637X/748/2/134}, 748, 134

\bibitem[\protect\citeauthoryear{Ghisellini, Celotti  \& Lazzati}{Ghisellini
  et~al.}{2000}]{Ghisellini2000}
Ghisellini G.,  Celotti A.,   Lazzati D.,  2000, \mn@doi [Monthly Notices of
  the Royal Astronomical Society] {10.1046/j.1365-8711.2000.03354.x}, 313

\bibitem[\protect\citeauthoryear{Giannios}{Giannios}{2008}]{Giannios2008}
Giannios D.,  2008, \mn@doi [åp] {10.1051/0004-6361:20079085}, 480, 305

\bibitem[\protect\citeauthoryear{Hascoët, Daigne  \& Mochkovitch}{Hascoët
  et~al.}{2012}]{Hascoet2012}
Hascoët R.,  Daigne F.,   Mochkovitch R.,  2012, \mn@doi [Astronomy and
  Astrophysics] {10.1051/0004-6361/201219339}, 542

\bibitem[\protect\citeauthoryear{Kobayashi, Piran  \& Sari}{Kobayashi
  et~al.}{1997}]{Kobayashi1997}
Kobayashi S.,  Piran T.,   Sari R.,  1997, \mn@doi [The Astrophysical Journal]
  {10.1086/512791}, 490, 92

\bibitem[\protect\citeauthoryear{Kowal, Falceta-Gonçalves, Lazarian  \&
  Vishniac}{Kowal et~al.}{2017}]{Kowal2017}
Kowal G.,  Falceta-Gonçalves D.~A.,  Lazarian A.,   Vishniac E.~T.,  2017,
  \mn@doi [The Astrophysical Journal] {10.3847/1538-4357/aa6001}, 838, 91

\bibitem[\protect\citeauthoryear{Kumar}{Kumar}{1999}]{Kumar1999}
Kumar P.,  1999, \mn@doi [The Astrophysical Journal] {10.1086/312265}, 523

\bibitem[\protect\citeauthoryear{Kumar \& McMahon}{Kumar \&
  McMahon}{2008}]{Kumar2008}
Kumar P.,  McMahon E.,  2008, \mn@doi [Monthly Notices of the Royal
  Astronomical Society] {10.1111/j.1365-2966.2007.12621.x}, 384

\bibitem[\protect\citeauthoryear{Lazarian, Zhang  \& Xu}{Lazarian
  et~al.}{2019}]{Lazarian2019}
Lazarian A.,  Zhang B.,   Xu S.,  2019, \mn@doi [The Astrophysical Journal]
  {10.3847/1538-4357/ab2b38}, 882

\bibitem[\protect\citeauthoryear{Liang, Lin, Lü, Lu, Zhang  \& Zhang}{Liang
  et~al.}{2015}]{Liang}
Liang E.-W.,  Lin T.-T.,  Lü J.,  Lu R.-J.,  Zhang J.,   Zhang B.,  2015,
  \mn@doi [The Astrophysical Journal] {10.1088/0004-637X/813/2/116}, 813, 116

\bibitem[\protect\citeauthoryear{Lu, Wei, Liang, Zhang, Lü, Lü, Lei  \&
  Zhang}{Lu et~al.}{2012}]{Lu2012b}
Lu R.~J.,  Wei J.~J.,  Liang E.~W.,  Zhang B.~B.,  Lü H.~J.,  Lü L.~Z.,  Lei
  W.~H.,   Zhang B.,  2012, \mn@doi [Astrophysical Journal]
  {10.1088/0004-637X/756/2/112}, 756

\bibitem[\protect\citeauthoryear{Lucas~Uhm, Zhang  \& Racusin}{Lucas~Uhm
  et~al.}{2018}]{LucasUhm2018}
Lucas~Uhm Z.,  Zhang B.,   Racusin J.,  2018, ] {10.3847/1538-4357/aaeb30}

\bibitem[\protect\citeauthoryear{Lyutikov \& Blandford}{Lyutikov \&
  Blandford}{2003}]{Lyutikov2003}
Lyutikov M.,  Blandford R.,  2003, arXiv e-prints, pp astro--ph/0312347

\bibitem[\protect\citeauthoryear{McKinney \& Uzdensky}{McKinney \&
  Uzdensky}{2012}]{McKinney2012}
McKinney J.~C.,  Uzdensky D.~A.,  2012, \mn@doi [{\textbackslash}mnras]
  {10.1111/j.1365-2966.2011.19721.x}, 419, 573

\bibitem[\protect\citeauthoryear{Mészáros \& Rees}{Mészáros \&
  Rees}{1997}]{Meszaros1997}
Mészáros P.,  Rees M.~J.,  1997, \mn@doi [{\textbackslash}apjl]
  {10.1086/310692}, 482, L29

\bibitem[\protect\citeauthoryear{Panaitescu, Spada  \& Mészáros}{Panaitescu
  et~al.}{1999}]{Panaitescu1999}
Panaitescu A.,  Spada M.,   Mészáros P.,  1999, \mn@doi [The Astrophysical
  Journal] {10.1086/312230}, 522

\bibitem[\protect\citeauthoryear{Racusin et~al.,}{Racusin
  et~al.}{2011}]{Racusin2011}
Racusin J.~L.,  et~al., 2011, \mn@doi [The Astrophysical Journal]
  {10.1088/0004-637X/738/2/138}, 738, 138

\bibitem[\protect\citeauthoryear{Rees \& Meszaros}{Rees \&
  Meszaros}{1994}]{Rees1994}
Rees M.~J.,  Meszaros P.,  1994, \mn@doi [The Astrophysical Journal]
  {10.1086/187446}, 430, L93

\bibitem[\protect\citeauthoryear{Rybicki \& Lightman}{Rybicki \&
  Lightman}{1985}]{Rybicki1985}
Rybicki G.~B.,  Lightman A.~P.,  1985, Radiative {Processes} in {Astrophysics},
  \mn@doi{10.1002/9783527618170.
}

\bibitem[\protect\citeauthoryear{Shao \& Gao}{Shao \& Gao}{2022}]{Shao2022}
Shao X.,  Gao H.,  2022, \mn@doi [The Astrophysical Journal]
  {10.3847/1538-4357/ac46a8}, 927, 173

\bibitem[\protect\citeauthoryear{Shen \& Zhang}{Shen \& Zhang}{2009}]{Shen2009}
Shen R.~F.,  Zhang B.,  2009, \mn@doi [Monthly Notices of the Royal
  Astronomical Society] {10.1111/j.1365-2966.2009.15212.x}, 398

\bibitem[\protect\citeauthoryear{Spitkovsky}{Spitkovsky}{2008}]{Spitkovsky2008}
Spitkovsky A.,  2008, \mn@doi [The Astrophysical Journal] {10.1086/590248},
  682, L5

\bibitem[\protect\citeauthoryear{Takamoto}{Takamoto}{2018}]{Takamoto2018}
Takamoto M.,  2018, \mn@doi [Monthly Notices of the Royal Astronomical Society]
  {10.1093/mnras/sty493}, 476, 4263

\bibitem[\protect\citeauthoryear{Takamoto, Inoue  \& Lazarian}{Takamoto
  et~al.}{2015}]{Takamoto2015}
Takamoto M.,  Inoue T.,   Lazarian A.,  2015, \mn@doi [The Astrophysical
  Journal] {10.1088/0004-637X/815/1/16}, 815, 16

\bibitem[\protect\citeauthoryear{Thompson}{Thompson}{1994}]{Thompson1994}
Thompson C.,  1994, \mn@doi [{\textbackslash}mnras] {10.1093/mnras/270.3.480},
  270, 480

\bibitem[\protect\citeauthoryear{Uhm \& Zhang}{Uhm \& Zhang}{2013}]{Uhm2013}
Uhm Z.~L.,  Zhang B.,  2013, ] {10.1038/nphys2932}

\bibitem[\protect\citeauthoryear{Uhm \& Zhang}{Uhm \& Zhang}{2015}]{Uhm2015}
Uhm Z.~L.,  Zhang B.,  2015, ] {10.3847/0004-637X/825/2/97}

\bibitem[\protect\citeauthoryear{Usov}{Usov}{1994}]{Usov1994}
Usov V.~V.,  1994, \mn@doi [{\textbackslash}mnras] {10.1093/mnras/267.4.1035},
  267, 1035

\bibitem[\protect\citeauthoryear{Xu, Huang, Geng, Wu, Li  \& Zhang}{Xu
  et~al.}{2023}]{Xu2023}
Xu F.,  Huang Y.-F.,  Geng J.-J.,  Wu X.-F.,  Li X.-J.,   Zhang Z.-B.,  2023,
  An analytic derivation of the empirical correlations of gamma-ray bursts,
  \url {http://arxiv.org/abs/2211.04727}

\bibitem[\protect\citeauthoryear{Yonetoku, Murakami, Nakamura, Yamazaki, Inoue
  \& Ioka}{Yonetoku et~al.}{2004}]{Yonetoku2004}
Yonetoku D.,  Murakami T.,  Nakamura T.,  Yamazaki R.,  Inoue A.~K.,   Ioka K.,
   2004, \mn@doi [The Astrophysical Journal] {10.1086/421285}, 609

\bibitem[\protect\citeauthoryear{Yonetoku, Murakami, Tsutsui, Nakamura,
  Morihara  \& Takahashi}{Yonetoku et~al.}{2010}]{Yonetoku2010}
Yonetoku D.,  Murakami T.,  Tsutsui R.,  Nakamura T.,  Morihara Y.,   Takahashi
  K.,  2010, \mn@doi [Publications of the Astronomical Society of Japan]
  {10.1093/pasj/62.6.1495}, 62, 1495

\bibitem[\protect\citeauthoryear{Zhang}{Zhang}{2018}]{Zhang2018book}
Zhang B.,  2018, The physics of gamma-ray bursts,
  \mn@doi{10.1017/9781139226530.
}

\bibitem[\protect\citeauthoryear{Zhang \& Mészáros}{Zhang \&
  Mészáros}{2002}]{Zhang2002}
Zhang B.,  Mészáros P.,  2002, \mn@doi [{\textbackslash}apj]
  {10.1086/344338}, 581, 1236

\bibitem[\protect\citeauthoryear{Zhang \& Pe'er}{Zhang \&
  Pe'er}{2009}]{Zhang2009}
Zhang B.,  Pe'er A.,  2009, \mn@doi [The Astrophysical Journal]
  {10.1088/0004-637X/700/2/L65}, 700, L65

\bibitem[\protect\citeauthoryear{Zhang \& Yan}{Zhang \& Yan}{2011}]{Zhang2011}
Zhang B.,  Yan H.,  2011, \mn@doi [Astrophysical Journal]
  {10.1088/0004-637X/726/2/90}, 726, 90

\bibitem[\protect\citeauthoryear{Zhang \& Zhang}{Zhang \&
  Zhang}{2014}]{Zhang2014}
Zhang B.,  Zhang B.,  2014, \mn@doi [Astrophysical Journal]
  {10.1088/0004-637X/782/2/92}, 782

\bibitem[\protect\citeauthoryear{Zhang et~al.,}{Zhang et~al.}{2007}]{Zhang2007}
Zhang B.,  et~al., 2007, \mn@doi [The Astrophysical Journal] {10.1086/510110},
  655, 989

\bibitem[\protect\citeauthoryear{Zitouni, Guessoum  \& Azzam}{Zitouni
  et~al.}{2016}]{Zitouni2016}
Zitouni H.,  Guessoum N.,   Azzam W.~J.,  2016, Revisiting the {Amati} and
  {Yonetoku} {Correlations} with {Swift} {GRBs}, \url
  {http://arxiv.org/abs/1611.05732}

\makeatother
\end{thebibliography}

\bsp	
\label{lastpage}
\end{document}